\documentstyle[floats,aps]{revtex}
\begin{document}
\draft

\catcode`\@=11 \catcode`\@=12
\twocolumn[\hsize\textwidth\columnwidth\hsize\csname@twocolumnfalse\endcsname
\title{Paired Hall States versus Unidirectional CDW in Tilted Field
 for $\nu=\frac{5}{2}$}
\author{Yue Yu, Shi-Jie Yang and Zhao-Bin Su}
\address{Institute of Theoretical Physics, Chinese Academy of Sciences, P. O. Box 2735,
Beijing 100080, China}

\maketitle
\begin{abstract}

We formulate the composite fermions in the presence of an in-plane magnetic field.
As the in-plane field increases, if we assume the state at $\nu=5/2$ turns into the
mixed state between the unidirectional charge density wave domains and paired
Hall state, we can phenomenologically fit the theoretically defined gap to the
experimental measured results. We explain the destruction of the paired Hall
states and then a phase transition from the paired Hall state to the unidirectional charge density wave from a symmetry point of view.

\end{abstract}

\pacs{PACS numbers: 73.20.Dx,73.40.Hm,73.40.Kp,73.50.Jt}]

There have appeared a bundle of enigmatic phenomena for half-filled Landau
levels since the discovery of the Hall metallic state for $\nu=1/2$ \cite
{jiang}and the Hall plateau for $\nu=5/2$ \cite{will}. Recently, a series of
experiments revealed a novel anisotropic electronic transport for
half-filled higher Landau levels \cite{ll1,du,sh,pan,ll}. It is widely
believed that the highly anisotropic transport is related to the formation
of the unidirectional charge density wave (UCDW) ground state, i.e., the
stripe phase, \cite{KFS,MC,HRY} as well as the quantum and smectic and nematic
phase induced by the fluctuations \cite{frad}.

While the composite fermion (CF) metallic states at $\nu=1/2$ was fairly
well-understood \cite{HLR}, the enigma at $\nu=5/2$ remains far to be revealed.
The Hall plateau at $\nu=5/2$ was explained as the appearance of the
ground state of a spin-singlet pair \cite{HaRe} whereas the $p$-wave BCS
paring of CF the spin-polarized or the Moore-Read(MR) Pfaffian wave function
\cite{MR} may be another possibility \cite{GWW}, which was recently suggested
to be favorable \cite{morf,RH}. Studies by Eisenstein et al \cite{Eis} in
the tilted field have shown that the plateau disappears if the tilt angle $%
\theta $ exceeds a critical value. The explanation of the experiments from
the point of view of the singlet-paring can be understood as a gain in
Zeeman energy \cite{Eis1} while how the tilted field violates the
spin-polarized paired Hall state is still in puzzle.

Recently, the experimental data showed by Pan et al \cite{pan} and Lilly et al
\cite{ll} shed light on this puzzling problem. These experiments revealed the
anisotropic transport property after the Hall plateau is destroyed by the tilt
field and the current favors to flow along the direction perpendicular to the
in-plane field. Two recent papers investigated the effect caused by the tilted field
\cite{St,Ju} but the phase transition mechanism from the quantum Hall state to
the stripe phase was not touched.

In this paper, we would try to search the physical mechanism causing this phase
transition . We explore that the many-body wave functions with the existence of the
in-plane field. It is found that the Laughlin-like states can be defined and then
the fractional quantum Hall effects (FQHE) and the Hall metallic state are not
disturbed by the tilted field for the lowest Landau level (LLL). However, there are
two kinds of instabilities of the CF Fermi surface for $\nu=5/2$, the CF paired
Hall state and the UCDW of the electrons. Before arriving the critical
tilt angle, the state may be the mixing between the paired Hall states
and the UCDW domains . It is seen that the paired Hall gap decreases as the tilt angle
increases. Moreover, it is shown that the theoretically defined gap
decreases as the total external magnetic field. Through a phenomenolgical
fit, our theoretical prediction for the gap function of the external magnetic field
agreeing with what Eisenstein et al presented in an earlier experiment \cite{Eis1}.
Thus, instead of the explanation based on a gain of the Zeeman energy, the
spin-polarized explanation of the experiment is established. Exceeding the
critical tilt angle, all UCDW domains connect together and form a global
UCDW.

From the symmetry point of view, the Laughlin-type states are only the eigen
state of the total magnetic translation operator whereas no longer those of
the relative one. This implies that the motion of the center of mass of
particles can not be separated from the relative motion. The lack of the
quantum numbers labeling the relative motion of the particles leads to the
pair was destroyed at a critical tilt angle for $\nu=5/2$.

We start from the problem of a single particle in a strong magnetic field
which is tilted an angle to the $x-y$ plane. An in-plane field in the $x$%
-direction violates the two-dimensional(2D) rotational symmetry. By
introducing a harmonic confining potential with the character frequency $
\Omega$ in the $z$-direction, the system is restricted to quasi-2D. To
regard Laughlin's states, one takes the symmetric gauge. The single particle
Hamiltonian can be diagonalized as $H_{{\rm s.p.}}=\hbar\omega_-\alpha^%
\dagger_\xi\alpha_\xi+\hbar\omega_ +\alpha^\dagger_z\alpha_z$. The
frequencies $\omega_\pm$ are given by \cite{St}
\begin{equation}
\omega_\pm^2=\frac{1}{2}(\tilde\Omega^2+\omega_c^2)\pm\frac{1}{2} \sqrt{(%
\tilde\Omega^2-\omega_c^2)^2+4|\tilde\omega|^2\tilde\Omega\omega_c},
\end{equation}
where $\tilde\omega=\omega_x(\omega_c/\tilde\Omega)^{1/2}$ and $\tilde\Omega%
^2=\Omega^2+\omega_x^2$; $\omega_x$ and $\omega_c$ are the cyclotron
frequencies corresponding to $B_x$ and $B_z$. $\alpha_\xi$ and $\alpha_z$
are the annihilation operators in the diagonal harmonic bases. Here we have
applied the unit $l_c=\sqrt{\hbar c/eB_z}=1$. In addition, there is a
conservation quantity, the square of the magnetic translation in 2D, $L_\xi=%
\tilde a_L^\dagger\tilde a_L$ with $\tilde a_L=\frac{1}{\sqrt{2}}%
(\partial_\xi+\frac{1}{2}\bar\xi)$ and $\tilde a^\dagger_L=\frac{1}{\sqrt{2}}
(-\partial_{\bar\xi}+\frac{1}{2}\xi)$. If $\theta$ tends to zero, $L_\xi$ is
corresponding to the angular momentum in 2D. To solve this single-particle
problem, we seek the ground state which is the engin function of $L_\xi$. It
is useful to make a coordinate rotation with $\xi\to\tilde\xi=\xi+b\bar\xi%
+cz^{\prime}$ and $\tilde z^{\prime}=z^{\prime}$ with $b$ and $c$ determined
by $[\alpha_\xi, \tilde\xi]= [\alpha_z, \tilde\xi]=0$. The ground state wave
functions are highly degenerate and of the form $\Psi_0(\tilde\xi,\tilde\xi
^*,\tilde z^{\prime})=f(\tilde\xi)e^g$ with $g(\tilde\xi,\tilde\xi ^*,\tilde %
z^{\prime})$ being a quadratic form of $\tilde\xi,\tilde\xi ^*,\tilde z%
^{\prime}$. The function $f(\tilde\xi)$ is an arbitrary function of $\tilde%
\xi$ and The coefficients of $g$ are determined by $\alpha_\xi e^g=\alpha_z
e^g=0$.

Notice that linear-independent wave functions $\tilde\xi^m e^g$
(m=0,1,2,...) are not the eigen functions of $L_\xi$. However, one can start
from those linear-independent wave functions to construct the common eigen
functions of $H_{{\rm s.p.}}$ and $L_\xi$, which read $f_m(\tilde\xi)e^g,$
with $f_m(\tilde\xi) =\sum_{m^{\prime}=0}^{M-1}f_{mm^{\prime}} \tilde\xi^m$
for $M$ being the number of Landau orbits. The coefficients $f_{mm^{\prime}}$
are dependent on the in-plane field and confined by $f_{mm}(0)=1$ and $%
f_{mm^{\prime}}(0)=0$ for $m\not{= }m^{\prime}$ if $\theta=0$. Those
degenerate ground state wave functions are orthogonal and with the eigen
value $m$ of $L_m$.

After solving the single-particle problem, we turn to the many-body ground
state wave function. To be enlightened by Laughlin's wave function for the
vanishing tilt angle, we postulate the many-body ground states for $\nu=1/%
\tilde\phi$ FQHE as $\Psi_0(\vec r_1,...,\vec r_N)=({\rm St}%
(f_0,...f_{N-1}))^{\tilde\phi} e^{\sum_i g_i} $ where $\vec r_i$ are the 3-d
position vectors and ${\rm St}$ is standing for the Slater determinant of $%
f_0(\tilde\xi_{P(1)}),..., f_{N-1}(\tilde\xi_{P(N)})$. This wave function
has the lowest eigen value of $L=\sum L_{\xi_i} $. For the small particle
system, the postulated wave function seems to be different from Laughlin's
wave function. However, if the numbers of both Landau orbits and particles
are very large, one can assume the function $f_m(\tilde\xi%
_i)=\sum_{m^{\prime}=0}^{N-1} f_{mm^{\prime}}\tilde\xi_i^{m^{\prime}}$ \cite
{com}. Thus, the Slater determinant is proportional to Vandermonde's
determinant and then
\begin{equation}
\Psi_0(\vec r_1,...,\vec r_N)\propto \Pi_{i<j} \tilde\xi_{ij}^{\tilde\phi%
}e^{\sum_i g_i},  \label{lau}
\end{equation}
which is exactly the Laughlin-Jastrow form for $\tilde \xi_{ij}=\tilde \xi%
_i- \tilde \xi_j$. For $\nu=1/2$, a similar boson-type Laughlin-Jastrow form
is also contained in the wave function. Thus, the composite particle picture
is still valid. In other words, we can make an anyon transformation
\begin{equation}
\Psi(\vec r_1,...,\vec r_N)=\Pi_{i<j}\biggl[\frac{\tilde\xi_{ij}} {|\tilde\xi%
_{ij}|}\biggr]^{\tilde\phi}\Phi(\vec r_1,...,\vec r_N).
\end{equation}
A statistical gauge field, therefore, is introduced,
\begin{equation}
{\tilde a}_\alpha(\vec r_i)=-\tilde\phi\sum_{j\not{=}i} \frac{\tilde\epsilon%
_{\alpha\beta} \tilde x^{\beta}_{ij}}{|\tilde\xi_{ij}|^2},~~ \tilde a_z(\vec %
r_i) =-\frac{\tilde\phi\tilde\Omega^{1/2}}{\omega_c^{1/2}} \sum_{j\not{=}i}%
\frac{c\tilde y_{ij}} {|\tilde \xi_{ij}|^2},
\end{equation}
where $\tilde\xi= \tilde x+i\tilde y$; $\tilde\epsilon_{12}=1+b$ and $\tilde%
\epsilon_{21}=-1+b$. This statistical gauge field gives an associated
statistical magnetic field $\vec b =\nabla\times \vec{\tilde a} $ with its
components
\begin{eqnarray}
b_z(\vec\xi_i)&=&-2\pi\tilde\phi\sum_{j\not{=}i}\delta^{(2)}(\xi_{ij}), \\
b_x(\vec\xi_i)&=&2\pi\tilde\phi(1+b)^{-1}(\tilde\Omega/\omega_c)^{1/2}
c\sum_{j\not{=}i}\delta^{(2)}(\xi_{ij}),  \label{bx} \\
b_y(\vec\xi_i)&=&0.  \nonumber
\end{eqnarray}
Hence, at the mean-field approximation, the perpendicular magnetic field can
be cancelled by taking the particle density $\rho(\vec \xi_i)=\sum_{j\not%
{=}i} \delta^{(2)}(\xi_{ij})$ as its average value $\bar\rho=\frac{\nu}{2\pi
l_c^2}$ for $\nu=1/\tilde\phi$. This gives the composite particles in the
zero perpendicular effective field. However, the in-plane field is enhanced
by the parallel statistical magnetic field. A convenient gauge choice is to
take $\bar{\tilde a}_y=B_xz$ and leave the rest to $\bar{\tilde a}_z$. Thus,
the residual vector potential under the mean-field approximation points to the $\hat z$-direction and does not affect the normal transport behaviors in the plane \cite{Mac}.

Now, turn to the second Landau level. We assume the mixing between Landau
levels can be neglected. Thus, the second Landau level can be treated as the
LLL except the interaction between particles is renormalized due to the
screening of the electrons in the LLL. In the absence of the in-plane field,
the enigmatic even-denominator Hall plateau was observed for $\nu=5/2$ \cite
{will}. An in-plane field, however, rapidly smashes the Hall plateau as the
magnetic field is tilted \cite{Eis}. Theoretically, as we have mentioned,
recent numerical simulations favor the spin-polarized ground state \cite
{morf,RH}. However, in the first glimpse, the $p$-wave paired Hall state can
only be affected by the tilted field gently. Therefore, a new explanation of
the tilted field experiments has to be constructed.

In fact, there are two kinds of the instabilities of the CF Fermi surface,
the CF paired Hall state and the UCDW of the electrons. The $p$-wave paired Hall state has an excitation gap  $\Delta(k)=\Delta_F(k/k_F)$ for $k>k_F$ and
$\Delta(k)=\Delta_F(k_F/k)$ for $k<k_F$ \cite{GWW}, where
$\Delta_F=\Delta(k_F)$. The energy difference between the paired state and
the CF Fermi sea can be obtained by using the standard expressions, i.e.,
\begin{equation}
E_{g_s}-E_{g_n}=2N(0)\int_0^{\epsilon_D}d\epsilon[\epsilon- \frac{%
2\epsilon^2+\Delta(k)^2}{2\sqrt{\epsilon^2+\Delta(k)^2}}],
\end{equation}
where $\epsilon_D$ is a cut-off.
For $\Delta_F\ll {\rm min}\{\epsilon_F,\epsilon_D\}$, one has
\begin{equation}
E_{g_s}-E_{g_n}=-\frac{1}{2}N(0)\Delta_F^2,
\end{equation}
i.e., to the zeroth order, the above energy difference is
cut-off independent.

As the magnetic field is tilted, the paired gap decreases. Furthermore,
eq.(\ref{bx}) means the local statistical in-plane field varies as the
local particle density if there is an in-plane field. Then, the density
fluctuation causes the effective in-plane field fluctuation. When one of CF's
in a pair of CF's is in the region where the effective in-plane field
is strongly enhanced the additional magnetic energy may break the pair into the normal
CF's. In terms of thermodynamics, the gap relates to the in-plane field by
\begin{equation}
\frac 12N(0)|\Delta _F(B_x)|^2=\frac 1{4\pi l}\int_{-l/2}^{l/2}dz%
\int_{B_x}^{B_p}\tilde{b}_xdB_x^{\prime },  \label{gap1}
\end{equation}
where $B_p=\sqrt{B_{{\rm tot,p}}^2-B_z^2}$ is defined by the paired Hall gap
at $\theta =0$; $l$ is the thickness of the layer; and $\tilde{b}_x$ is the
distribution of the magnetization in the $\hat{z}$-direction. The boundary
condition is $\tilde{b}_x(l/2)=\tilde{b}_x(-l/2)=\bar{b}_x$ which is the
average of $b_x$ in the plane, i.e., the mean-field value of $b_x$. (This is non-linear
to $B_x$. However, this non-linearity is very small since $
\theta $ is small. We can still assume $\bar b_x=\gamma B_x$.) If we thought this
is the gap observed in the experiment, $B_p$ would equal to $B_{x,c}$ and (\ref{gap1})
could be rewritten as
\begin{equation}
\frac 12N(0)|\Delta _F(B_{{\rm tot}})|^2=\frac{\gamma \Lambda }{8\pi }%
(B_{x,c}^2+B_z^2-B_{{\rm tot}}^2),  \label{gap2}
\end{equation}
where $B_{{\rm tot}}^2=B_z^2+B_x^2$; $\Lambda $ is a factor related to the
enhancement of the critical field due to the thin film in which the
penetration depth of the in-plane magnetic field is larger than $l
$. However, the gap function depending quadratically on $B_{{\rm tot}}$ in (
\ref{gap2}) is qualitatively different from that in the experiment by
Eisenstein et al \cite{Eis1}. This means that the explanation to the
experiment from the mixed state like the superconducting thin film may not
work. Moreover, if such a mixed state worked, one would observe an
anisotropic transport with the easy direction along the $B_x$-direction
. This is just opposite to the experimental observation.
The reason for the CF Fermi liquid not working is that the CF Fermi sea may
not a good variational ground state. It has been known that the ground state
may be the UCDW perpendicular to the in-plane field if $\theta >\theta _c$,
which has been supported by recent numerical calculations \cite{Ju}. The experiments
have also supplied the evidence that the stripe favor such an orientation \cite
{ll,pan}. Thus, when the effective in-plane magnetic field exceed a critical value in a domain, the CF's pairs are broken and the normal
CFs energetically favor to change back to the electrons. The electrons forms a
UCDW domain. Because $b_x$ increases as $B_x$, the number and size of UCDW domains
increase. After all, these domains all connection together at the critical
tilted angle. The question is if the cohesive energy for the electrons in a
very small domain when $B_x$ is very small is still negative?
To answer this question, we note that unlike in the real zero- or
one-dimensional system, all electrons in the bulk they are not confined in a
small region by any barrier or potential and the single-electron wave
function is still taken its two-dimensional version. Although the charging effect
is not important for CF's due to the neutrality of the bulk CF excitation, it comes
over when CFs move into these domains because of the possible UCDW instability.
According to ref.\cite{KFS}, the cohesive energy is not dependent on the length of the stripes if the single-particle wave function takes its two-dimensional version.
Thus, the cohesive energy $E_{\rm coh}$ for the electrons in the small domain
can be calculated in the Hartree-Fock approximation as done in
literature \cite{KFS,MC,St,Ju}. Several particular values are listed in table 1.

\hspace{0.3cm}

\begin{tabular}{ccccc}
\hline
$\theta$ & 0$^0$ & 7.2$^0$ & 14.4$^0$ & 20.4$^0$ \\ \hline
~~E$_{\rm coh}(K)$~~ & ~~ -6.251~~ & ~~ -6.252~~ & ~~ -6.253~~ & ~~-6.255~~ \\
\hline
\end{tabular}

\hspace{0.2cm}

{\small Table 1 The typical values of the cohesive energy versus $B_{\rm tot}$.
The in-plane field is chosen to be perpendicular to the stripes \cite{comm}.}

We have given a mixed state picture between the paired Hall state and the UCDW.
To exactly evaluate the microscopic parameters $\gamma$ and $\Lambda$
are difficult because their bare mean field value have to be strongly
renormalized by the gauge fluctuation. Here we take a phenomenological method to
deal with. The real energy gap of the Fermi momentum $k_F$, $
\Delta (B_{{\rm tot}})$, which is assumed being observed in the experiment,
is given by
\begin{equation}
\Delta (B_{{\rm tot}})=\Delta _F(B_{{\rm tot}})-|E_{\rm coh}(B_{{\rm tot}})|,
\label{gap3}
\end{equation}
according to the mixed state picture.
The parameters $\Delta _F(B_z)$ (or the combination of the parameters $
\gamma\Lambda/2\pi N(0)$) and $B_{{\rm tot,p}}$ can be determined by
the experimental value of $\Delta (B_z)$ and the vanishing gap at $B_{{\rm
tot}}=B_{{\rm tot,c}}$, i.e., $\Delta (B_{{\rm tot,c}})=0$ with $B_{{\rm
tot,c}}$ read out from the experimental data. However, the experimentally
measured gap at $\theta =0$ is 0.11 K which is one order differing from the
theoretical calculation ($\sim 2$K). This discrepancy may come from the
residual disorder in the sample \cite{Tn}. A first order approximation to
deal with the disorder is simply to minus a constant $\Gamma $ in the right
hand side of (\ref{gap3}) \cite{Eis2} such that $\Delta (0)=0.11$K. The
energy gap $\Delta $ versus $B_{\rm tot}$, then, can be calculated, which is
plotted in Fig. 1. A fit to the experiment data is shown. It is emphasized
that instead of the exactly linear energy gain from the Zeeman energy in the
earlier spin unpolarized model, our result is not a rigorously straight line,
which seems to be more reasonable in comparing with the experiment data. In
fact, the curve drawn in Fig.1 comes from the very beginning part of the
descent parabola of $B_{{\rm tot}}$ pluses the cohesive energy which
slightly decreases as $B_{{\rm tot}}$ increases.

We can understand the destruction of the paired Hall gap from symmetry
point of view. Recall the MR Pfaffian wave functions \cite{MR}, the CF pair
has the eigen value $l=-1$ of the relative angular momentum, i.e., the $p$
-wave pairing. With the tilted field, we can also separate the conservation
operator $L=\sum L_{\xi_i}$ into two parts one of which is corresponding to
the motion of the center of mass in $\tilde\xi$-plane and another to the
relative motion. Defining $\Xi=\frac{1}{N}\sum_i\tilde\xi_i$ and $\delta_i=
\frac{1}{\sqrt{2}}(\tilde\xi_{i+1}-\tilde\xi_i)$, $L$ is divided into $%
L=L(\Xi)+L(\delta)$. It is easy to see that $\sum_i g_i$ can also be written
as $g(\Xi)+g(\delta)$. However, Slater determinant does not consist of $%
f_m(\Xi)$ and $f_m(\delta)$. This means that the wave function of the ground
state is neither the eigen wave function of $L(\Xi)$ nor of $L(\delta)$. For
example, although Laughlin's states (\ref{lau}) can be written into the form
of $F(\delta)e^{g(\Xi)}$ with $F(\delta)=\Pi_{i<j}(\tilde\xi_{ij}) ^{\tilde%
\phi} e^{g(\delta)}$, neither $F(\delta)$ nor $e^{g(\Xi)}$ is the eigen
state of $L(\delta)$ or $L(\Xi)$, respectively. Thereby, we see that the
relative motion of the particles is not independent of the motion of the
center of mass due to the tilted field. A single pair can not be labeled by
a well-defined quantum number. Furthermore, the paired Hall states,
especially the MR Pfaffian wave function are not eigen states of $L$ (then $%
L(\Xi)$ and $L(\delta)$).Certainly, in a very weak in-plane field, the
effect of the field can be thought as a perturbation, and pairs may still be
identified. As $\theta$ increasing, the mixing between the motion of the
center of mass and the relative motion is stronger. Finally, a phase
transition from the paired Hall to a compressible liquid takes place. Since
the CFs are disintegrated at $B_{{\rm tot,c}}$ and the level crossing
from the paired Hall state to the UCDW happens in the ground state. The
effective magnetic field contributing to the entropy has a sudden jump from $%
B_x^*=B_x- \bar b_x$ to $B_x$. Therefore, this phase transition is the first
order one.

In conclusions, we have constructed the CF when the magnetic field is
tilted. While Laughlin-like states are well-defined in the LLL, the $p$-wave
paired Hall state can be unstable because of the residual in-plane field. A
competition between the instabilities of the CF Fermi surface to the
formation of the CDW and the paired Hall state leads to the mixed state we
considered here. Finally, the UCDW takes energetically over the paired Hall
state, which transforms the incompressible state to the compressible state.
One found that theoretically defined gap versus $B_{\rm tot}$ is well fitted to
the experimental results.

This work was supported in part by the NSF of China.

Fig. 1 The paired gap $\Delta$ vs the total field $B_{\rm tot}$. The triangles
are the experimental data from ref. \cite{Eis1} and the solid line is the
theoretical result. $B_{\rm tot,c}\approx 4.05$T can be read out from the
extrapolation. $B_z=3.75$T.


\begin{references}
\bibitem{com}  This is exact if $\nu =1$, i.e., $N=M$. For fractional
filling of the Landau level, one can assume a truncation such that $%
f_{mm^{\prime }}=0$ if $m\geq N$ or $m^{\prime }\geq N$.
\bibitem{comm}  Even this choice is in the wrong direction with respect to
the experiment, this is still does not affect the fitting result in Fig. 1
too serious because the anisotropy of $E_{\rm coh}$ is only of several mK.
\end{references}
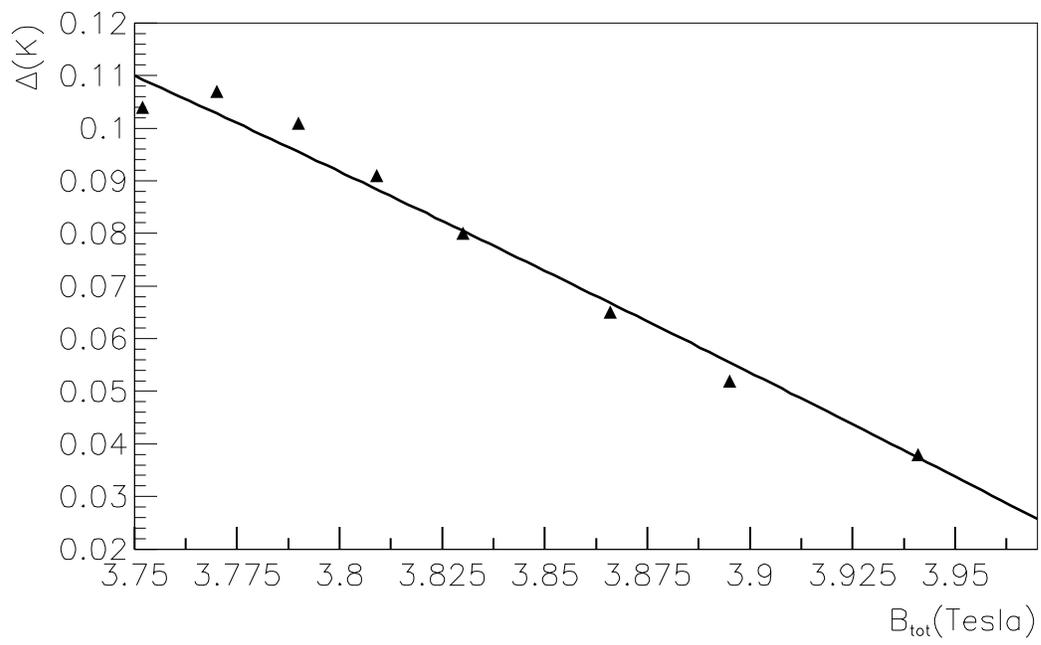
\end{document}